\documentclass[final,5p,times,twocolumn]{elsarticle}
\usepackage{amssymb}
\usepackage{graphicx}
\usepackage{dcolumn}
\usepackage{amsmath}

\begin{document}

\begin{frontmatter}

	\title{Topological phase transitions in strongly correlated systems: application to Co$_3$Sn$_2$S$_2$}
\author{V. Yu. Irkhin}

\ead{Valentin.Irkhin@imp.uran.ru}
\author{Yu. N. Skryabin}
\address
{M. N. Mikheev Institute of Metal Physics, 620108 Ekaterinburg, Russia
}

\begin{abstract}
	The topological  transition in the strongly correlated half-metallic ferromagnetic compound Co$_3$Sn$_2$S$_2$ from Weyl semimetal (including chiral massless fermions) to a non-magnetic state is treated. This transition goes with  a change in topological invariant, and is accompanied by a non-topological transition from saturated  ferromagnetic to paramagnetic state, the minority Fermi surface being transformed from ghost (hidden) to real. A corresponding description is given in terms of slave fermion representation for the effective narrow-band Hubbard model. The system Co$_3$Sn$_2$S$_2$ provides a bright example of coexistence of non-trivial topology and strong low-dimensional ferromagnetism.
	A comparison is performed  with  other compounds where frustrations result in formation of a correlated paramagnetic state.
\end{abstract}


\end{frontmatter}

\section{Introduction}
\begin{figure}[h]
	\includegraphics[width=0.45\textwidth]{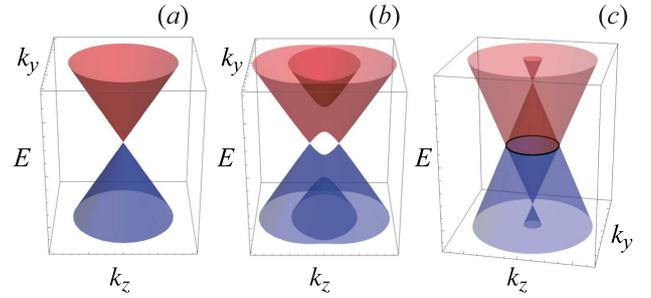}
	\caption{Schematic energy spectra  for the Dirac semimetal (a), Weyl semimetal (b), and nodal line  semimetal (c), according to Refs.\cite{Koshino,Armitage}.}
	\label{1}
\end{figure}

Recently, the layered kagome lattice compound Co$_3 $Sn$_2 $S$ _2 $ has been a subject of numerous experimental \cite{Wang,Xu,Liu,Liu2,Jiao,Zhou,Muechler,Tanaka,Yin,Xu1,Guguchia.NatComm,Liu3,Shin} and theoretical  \cite{Yanagi,Savrasov} investigations. In particular, giant topological anomalous Hall effect was observed in this system \cite{Liu,Zhou,Tanaka}. Its electronic structure  contains Weyl points, Fermi arcs and nodal rings (Fig. \ref{1}), which play an important role in the anomalous Hall effect  \cite{Liu,Tanaka,Yanagi}. 

Single-crystal experimental data on the Co$_3$In$_x$Sn$_{2-x}$S$_2$  kagome system \cite{Kassem} show that these systems have an almost two-dimensional itinerant magnetism and a chiral spin state; 
in addition, a strongly correlated state with a high electronic heat capacity is formed. The important role of correlations is confirmed by a considerable enhancement of $\gamma T$-linear specific heat even in the ferromagnetic phase  \cite{Schnelle,Kassem}, especially at approaching the magnetic-nonmagnetic critical point somewhat below $x=1$. At $x=1$, $\gamma$ vanishes, but strongly increases with further increasing $x$ \cite{Kassem}.   

The ferromagnetism in  Co$_3 $Sn$_2 $S$ _2 $ breaks time-reversal $T$-symmetry and is necessary for the existence of topological Weyl points. 
The strongly correlated metallic state in this system emerges as a result of magnetic fluctuations. Above $T_C$, intrinsic magnetic field disappears, the Weyl points annihilate and the Dirac points acquire a gap. This restores $T$-symmetry and eliminates the topological behavior.

\begin{figure}[h]
	\includegraphics[width=0.45\textwidth]{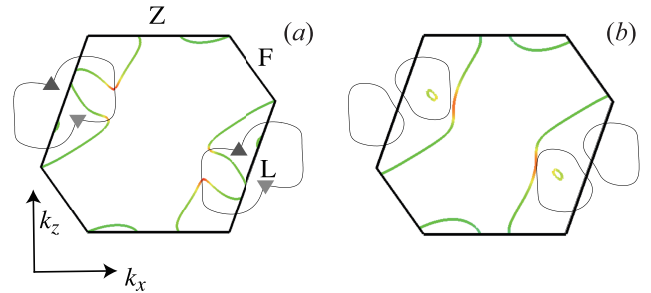}
	\caption{Schematic Fermi surfaces (solid green lines) for  Co$_3 $Sn$_2 $S$ _2$ in the $k_x-k_z$ plane at $k_y = 0$  according to  Ref.\cite{Yanagi}. The thin black solid line shows the nodal lines in the absence of spin-orbit coupling for (a) $x = 0.2$ and (b) $x = 0.4$. The upper and lower triangles on the nodal lines in (a) stand for the Weyl points with topological charges $+1$ and $-1$ in the presence of spin-orbit coupling.}
	\label{2}
\end{figure}

A  similar, but quantum transition occurs with disappearance of ferromagnetism in the Co$_3$In$_x$Sn$_{2-x}$S$_2$ system at the hole doping \cite{Zhou,Yanagi}. The  doping 
shifts the Weyl nodes away from the Fermi level. For small doping, the nodal rings are located around the Fermi energy,  
and for $x \sim 0.2$, the nodal lines surrounding the L point in the Brillouin zone  cross the Fermi surfaces (Fig. \ref{2}).  
With further increasing $x$, the nodal lines are split into two rings as with the annihilation of Weyl points  in
the presence of the spin-orbit coupling (the role of the latter is discussed in detail in Ref. \cite{Liu3}). 
For $x >0.6$, the nodal lines are located far from the Fermi level, resulting in the small Berry
curvature on the whole Fermi surfaces \cite{Yanagi}. At $x=1$ the system becomes insulating \cite{Zhou}.
According to \cite{Kassem}, this anomalous nonmetallic state may originate from the Fermi energy tuning through a Dirac point.

In the present work we treat the model picture of correlated half-metallic ferromagnetism in Co$_3 $Sn$_2 $S$ _2 $ 
and provide a description of these transitions within the topological classification \cite{Volovik,Volovik1,Volovik2}.

\section{Half-metallic ferromagnetism and Weyl points in the effective Hubbard model}

The half-metallic ferromagnetism of  Co$_3 $Sn$_2 $S$ _2 $ occurs in the partially filled Co 3$d_{x^2 - y^2}$ band which crosses the Fermi level.  The itinerant magnetism  cannot be fully described by a simple one-electron approach including band splitting.
In particular, the persistence of local moments from in the Curie Weiss susceptibility suggests a  many-body correlated behavior \cite{Savrasov}.
The associated moment of 1$\mu_B$ is spread over three Co atoms, in agreement with the 0.33$\mu_B$ per Co magnetic moment from first-principle calculations and  the experimental moment which is slightly less than 1$\mu_B$/f.u. The configuration with  five electrons and one hole results in a spin-1/2 ferromagnetic state. This enables one to formulate a local Hubbard model for the three Co atom cluster \cite{Savrasov}.

According to \cite{Savrasov}, across the magnetic transition,  Co$_3 $Sn$_2 $S$ _2 $ evolves from a Mott ferromagnet to a correlated metallic state. In fact, the  ``Mott ferromagnet'' is a half-metallic ferromagnetic state, so that we have a partial Mott transition in the minority spin subband. The physical reason of half-metallic ferromagnetism in a wide doping region is the presence of Van Hove singularity at the Fermi level, which is connected with quasi-two-dimensional kagome structure (see Fig.2 in Ref. \cite{Yanagi}). First-principle  calculations \cite{Yanagi} and experimental data \cite{Zhou,Kassem} demonstrate that the doping in  Co$_3$In$_x$Sn$_{2-x}$S$_2$ system preserves half-metallic ferromagnetism with a linearly decreasing magnetic moment. 

The picture of half-metallic ferromagnetism can be qualitatively described by the simplest narrow-band Hubbard model with large on-site repulsion $U$. In this model, doubly occupied states (doubles) are absent owing to the Hubbard splitting, but states with both spin projections are still present. Thus the situation is different from the Stoner model where spin splitting becomes infinitely large.
The corresponding Hamiltonian  reads
\begin{equation}  \label{eq:original_H}
	\mathcal{H}_{\mathrm{H}} = \sum_{ij\sigma}
	t_{ij}\tilde{c}^\dag_{i\sigma}\tilde{c}^{}_{j\sigma}
\end{equation}
where 
\begin{equation} 
	\tilde{c}^{\dagger}_{i\sigma}=X_i(0,\sigma)=|i0 \rangle \langle i \sigma|
\end{equation} 
are the Hubbard projection operators describing motion of holes in the correlated state on the background of magnetic moments.
We can use the slave fermion representation 
\begin{equation}
	X_{i}(0,\sigma )=f_{i}^{\dagger}b_{i\sigma },\, X_{i}(+,- )= b^\dag_{i\uparrow }b_{i\downarrow },  \label{eq:6.31}
\end{equation}%
where  $f_{i}$ are fermions and $b_{i\sigma }$ are  Schwinger boson operators (cf. \cite{Kane,IS}), so that
\begin{equation}
	\sum_{\sigma}b^\dag_{i\sigma }b_{i\sigma }+f_{i}^{\dagger}f_{i}=1.  \label{eq:6.122}
\end{equation}%
In the saturated ferromagnetic state the $b_{i\uparrow }$ boson is condensed, $\langle b_{i\uparrow } \rangle \simeq 1$, and $b_{i\downarrow }$ becomes magnon annihilation operator.

Consider the hole Green's functions for a saturated ferromagnet in the representation (\ref{eq:6.31}). 
\begin{equation}
	G_{\mathbf{k}\sigma}(E)=\langle \langle X_{\mathbf{k}}(\sigma,0)|X_{\mathbf{-k}}(0,\sigma)\rangle \rangle _{E}.
\end{equation}
The spin-up (majority) states propagate freely on the background of  strong ferromagnetic ordering, 
\begin{equation}
	G_{\mathbf{k}\uparrow }(E)=(E-t_{\mathbf{k}})^{-1}
\end{equation}
with $t _{\mathbf{k}}$  the bare band spectrum.
For our system, these possess an exotic spectrum of chiral Weyl fermions in the internal field owing to ferromagnetic ordering of local moments.  This just leads to unusual topological properties.

The spin down (minority)  Green's function in the leading approximation is obtained as a convolution of the Green's functions for free fermions and bosons, so that  
\begin{equation}
	G_{\mathbf{k}\downarrow }(E)=\sum_{\mathbf{q}}\frac{N_B(\omega_{\mathbf{q}%
		})+f(t_{\mathbf{k}+\mathbf{q}})}{E-t _{\mathbf{k}-\mathbf{q}}+\omega _{\mathbf{q}}}  \label{eq:I.779}
\end{equation}
where $N_B(\omega)$ and $f(E)$ are the Bose and Fermi functions,  $\omega_{\mathbf{q}}$ is the magnon spectrum. 
Similar results for  a Hubbard ferromagnet  were obtained earlier in
the many-electron representation of X-operators \cite{FTT,JPCM}, the
analogy with Anderson's spinons \cite{633a} being discussed. 

\begin{figure}[h]
	\includegraphics[width=0.45\textwidth]{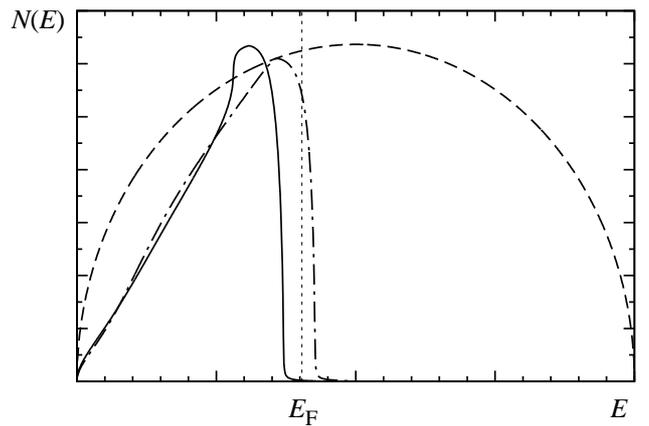}
	\caption{Schematic picture of the density of states in the narrow-band Hubbard model illustrating the doping-induced Lifshitz transition, the data are from Ref. \cite{Zarubin}. Dashed line is the bare (and also spin-up) density of states, solid and dot-dashed lines are the spin-down densities of states for the half-metallic and usual ferromagnetic states. 
	}
	\label{3}
\end{figure}

The Green's function (\ref{eq:I.779}) 
has a purely non-quasiparticle nature. Because of very weak $%
\mathbf{k}$-dependence of the corresponding distribution function the
non-quasiparticle (incoherent) states  do not carry current.
As demonstrate more exact calculations \cite{JPCM}, at small doping 
the spin down Green's function (\ref{eq:6.31}) has no poles
above the Fermi level of holes $E_F$, so that the above conclusions are not changed.
However, with increasing doping, it can acquire a spin-polaron pole above $E_F$,
and the half-metallic ferromagnetism is destroyed (Fig.\ref{3}).

Despite such unusual properties, the minority states cannot be fully neglected. In fact,  
their number is equal to the number of majority states $n_0$ owing to  the sum rule
\begin{equation}
	\sum_{\mathbf{k}}\langle X_{\mathbf{-k}}(0,\sigma)X_{\mathbf{k}}(\sigma,0)\rangle= \langle X_i(0,0)\rangle=n_0
\end{equation}
for both projections $\sigma$ (which is satisfied in the approximation (\ref{eq:I.779})), so that the current carriers (Hubbard's holes) are  in a sense spinless (see Fig.3: the corresponding areas below the Fermi level determining the occupations numbers are nearly equal).
The minority states in the ground state are absent at the Fermi surface and occur there at finite temperatures owing to thermal magnon excitations. The physics does not qualitatively change in the case of finite Hubbard $U$, since the doubles are automatically eliminated in the saturated half-metallic state \cite{JPCM}.

The description of the transition to the half-metallic state (vanishing of the
quasiparticle residue for one spin projection) is similar to that of the Mott transition in the
paramagnetic Hubbard model \cite{Senthil}. Here, the Fermi surface becomes ghost (hidden) after 
the metal-insulator transition in the insulating (Mott) phase, and the fractionalization of 
electron states occurs, including spin-charge separation into a neutral fermion (spinon) and charged boson (holon). 
Our situation can be described as a partial Mott transition in the minority spin subband, cf. the
discussion of the orbital-selective Mott transition  \cite{Vojta}.

The electron states in a strongly correlated system need not to have purely quasiparticle nature. 
They can be described by both poles and branch cuts of the Green's function. For minority states, 
the quasiparticle residue $Z$  is completely suppressed,   
which means formation of the energy gap and takes place, e.g., for the Mott transition \cite{Volovik}.
The violation of the standard Fermi-liquid picture can be described in terms
of the formation of the Luttinger surface which is the surface of zeros of the electron Green's function.

The Lifshitz transitions with vanishing quasiparticle poles can be viewed as quantum phase transitions  with a  change of the topology of the Fermi surface, but without symmetry breaking.  Indeed, the Fermi surface itself  is  the singularity in the Green's function, which is characterized by topological invariant $N_1$ and topologically protected: it is the vortex line in the  frequency-momentum space \cite{Volovik,Volovik1}.
In the gapped phase, usual Fermi surface does not exist, but its topology  is preserved if we take into account the Luttinger contribution.
Then the Luttinger theorem (the conservation of the volume enclosed by the Fermi surface) is still valid \cite{Volovik}.
Thus the Fermi surface becomes  ghost (hidden) in the Mott phase for both spin projections and in our half-metallic situation  for minority states, since the Fermi level lies in the corresponding gap.

On the contrary, the transition with disappearance of the Weyl points is essentially topological:  topological  invariants are changed. 
In the Weyl semimetal phase, the Weyl points have topological charges $N_3= +1$ and $-1$ and annihilate in the critical  Dirac semimetal. Further on, in the normal paramagnetic state the topology owing to the Berry curvature vanishes. Thus the conservation law for the topological charge \cite{Volovik} is fulfilled.
In the insulator case, we have a transition from topological to normal  insulator with restoring time-reversal symmetry. A still more complicated situation occurs in the case of Chern insulators with a change of the Chern number \cite{Hasan,Muechler,My}.

At the hole doping in the Co$_3$In$_x$Sn$_{2-x}$S$_2$ system, suppression of ferromagnetism is also accompanied with decreasing the Berry curvature, and in the paramagnetic phase both the Weyl points and nodal rings vanish  \cite{Yanagi}.

\section{Discussion}

As stated in Ref. \cite{Savrasov}, the topological properties and strong correlations in  Co$_3$Sn$_2 $S$_2$  are intricately linked, so that one cannot be adequately considered without the other.
The scanning tunnelling microscopy data in  Co$_3$Sn$_2 $S$_2$ demonstrated a pronounced peak at the Fermi level, which was identified as arising from the kinetically frustrated kagome flat band \cite{Yin}. The occurrence of an electronic band connecting the two Weyl cones and flattened by electronic correlations was demonstrated in Ref. \cite{Xu1},  the Coulomb-interaction strength being estimated as $U \sim 4$ eV.

The electron correlations in pyrochlore iridates $R_2$Ir$_2$O$_7$ were discussed by first-principle calculations in Ref. \cite{Wan}.
Depending on the strength of correlations $U$, a Mott insulating phase with a magnetic
structure or Weyl semimetal phase with  Weyl points  at the Fermi energy were treated. A variety of phases ranging from normal metal at small $U$ to Weyl semimetal at intermediate $U \sim$ 1.5 eV and Mott insulating phase at $U$ above 2 eV with non-collinear magnetic all-in/all-out ordering were predicted.
The giant anomalous Hall effect is observed in Nd$_2$Mo$_2$O$_7$, a pyrochlore
ferromagnet with geometrically frustrated lattice structure, is mostly due to
the spin chirality and the associated Berry phase originating from the Mo spin
tilting \cite{Taguchi}. However, the situation in pyrochlores with strong spin-orbit coupling differs from  half-metallic ferromagnetism.

We have demonstrated that in the half-metallic ferromagnetic state Hubbard correlations do not result in narrowing of bare bands for majority states, but in the paramagnetic state the situation changes: we come to the regime of narrow correlated bands for both spin projections. These may be characterized either by strongly renormalized quasiparticle residue, or even by a non-Fermi-liquid (e.g., marginal Fermi-liquid \cite{Senthil}) behavior. Besides absence of  $T$-breaking internal magnetic field in the paramagnetic phase, this can be important for vanishing of topological effects.

Since magnetic frustration effects in the kagome lattice  should play a role, the magnetic structure of Co$_3$Sn$_2 $S$_2$ is to be discussed in more detail. 
According to  \cite{Shin}, at finite temperatures this includes the out-of-plane ferromagnetism, in-plane antiferromagnetism, and hidden phases. A metastable magnetic phase  exists in some interval $T_{com}<T<T_C$. The corresponding values of transition temperatures are $T_C = 182$ K, $T_N = 177$ K, and $T_{com} = 150$ K.
The out-of-plane magnetization of Co$_3$Sn$_2 $S$_{2-x}$Se${_x}$ demonstrates a first-order phase transition, which may again indicate strong half-metallic magnetism. This may also important for a combined description of the non-topological ferromagnetic and  topological transitions. 
A possibility of a chiral spin state is discussed in \cite{Kassem}.

From the $\mu$SR measurements, the ground ferromagnetic state with moment along c axis exists in Co$_3$Sn$_2 $S$_2$ below $T^*_C$ = 90 K, and a coexistence of the ferromagnetic  order and an in-plane 120$^{\circ}$ antiferromagnetic order was proposed  at $T>T^*_C$  \cite{Guguchia.NatComm} (see, however, Ref.\cite{Soh}).
The  antiferromagnetic volume  fraction  grows  with increasing temperature and dominates around 170 K, before it disappears at $T_{C2}=172$ K. Above $T_{C2}$, the sample has the small volume fraction with the out-of-plane ferromagnetic order, the rest of the volume being occupied by the paramagnetic state up to  $T_{C1}=177$ K.  
The temperature-dependent magnetic fraction  shows a rather sharp transition between the paramagnetic and magnetic states with the coexistence of their regions in the interval $T_{C2}<T<T_{C1}$.

Thus, although frustrations in Co$_3$Sn$_2 $S$_2$ seem to be important, they turn out to be insufficient for formation of a paramagnetic spin-liquid-like state. The situation seems to be different for the correlated kagome systems  YCr$_6$Ge$_6$ \cite{Yang} and Na$_{2/3}$CoO$_2$ \cite{Alloul} without magnetic ordering.



Besides pyrochlores,  comparison can be made with other three-dimensional systems.
Two topological Weyl cones with band crossing points were identified around the X point for the Heusler alloy Co$_2$MnGe, 
which can induce large anomalous Hall  effect owing to the Berry flux in the half-metallic ferromagnet structure \cite{Kono}.


To conclude, Co$_3$Sn$_2 $S$_2$  provides a bright example of coexistence of non-trivial topology and half-metallic ferromagnetism in a quasi-two-dimensional system. Both these factors are important for non-usual phase transitions and anomalies of electronic properties, including giant anomalous Hall effect.

The authors are grateful to A. V. Zarubin for the help in preparing the manuscript.
The research was carried out within the state assignment of FASO of Russia (theme ``Flux'' No AAAA-A18-118020190112-8 and theme ``Quantum'' No. AAAA-A18-118020190095-4). 

{}

\begin{thebibliography}{99}
	
	
	\bibitem{Wang}
	Q. Wang, Y. Xu, R. Lou, Zh. Liu, M. Li, Y. Huang, D. Shen, H. Weng, Sh. Wang, H. Lei,
	Nature Comm. \textbf{9,} 3681 (2018).
	
	\bibitem{Xu}
	Q. Xu, E. Liu, W. Shi, L. Muechler, J. Gayles, C. Felser, and Y. Sun, 
	Phys. Rev. B \textbf{97}, 1 (2018).
	
	
	\bibitem{Liu}
	E. Liu, Y. Sun, N. Kumar, L. Muechler, A. Sun, L. Jiao, Sh.-Y. Yang, D. Liu, A. Liang, Q. Xu, J. Kroder, V. Seuss, H. Borrmann, Ch. Shekhar, Zh. Wang, Ch. Xi, W. Wang, W. Schnelle, S. Wirth, Y. Chen, S. T. B. Goennenwein and C. Felser,
	Nature Phys.  \textbf{14}, 1125 (2018). 
	
	
	\bibitem{Liu2}
	D. F. Liu, A. J. Liang, E. K. Liu, Q. N. Xu, Y. W. Li, C. Chen, D. Pei, W. J. Shi, S. K.
	Mo, P. Dudin, T. Kim, C. Cacho, G. Li, Y. Sun, L. X. Yang, Z. K. Liu, S. S. P. Parkin, C.
	Felser, and Y. L. Chen, 
	Science \textbf{365}, 1282 (2019).
	
	
	\bibitem{Jiao}
	L. Jiao, Q. Xu, Y. Cheon, Y. Sun, C. Felser, E. Liu, and S. Wirth, 
	Phys. Rev. B \textbf{99}, 1 (2019).
	
	\bibitem{Zhou}
	H. Zhou, G. Chang, G. Wang, X. Gui, X. Xu, J.-X. Yin, Z. Guguchia, S. S. Zhang, T.-R. Chang, H. Lin, W. Xie, M. Z. Hasan, Sh. Jia,
	Phys. Rev. B\textbf{101}, 125121 (2020).
	
	
	\bibitem{Muechler}
	L. Muechler, E. Liu, J. Gayles, Q. Xu, C. Felser, Y. Sun, Phys. Rev. B \textbf{101}, 115106 (2020).
	
	
	\bibitem{Tanaka}
	M. Tanaka, Y. Fujishiro, M. Mogi, Y. Kaneko, T. Yokosawa, N. Kanazawa, S. Minami, T. Koretsune, R. Arita, S. Tarucha, M. Yamamoto, and Y. Tokura,
	Nano Lett.  \textbf{20},  7476 (2020).
	
	
	\bibitem{Yin}
	J.-X. Yin, S. S. Zhang, G. Chang, Q. Wang, S. S. Tsirkin, Z. Guguchia, B. Lian, H. Zhou, K. Jiang, I. Belopolski, N. Shumiya, D. Multer, M. Litskevich, T. A. Cochran, H. Lin, Z. Wang, T. Neupert, Sh. Jia, H. Lei and M. Z. Hasan, Nature Physics \textbf{15}, 443 (2019).
	
	\bibitem{Xu1}	
	Y. Xu, J. Zhao, C. Yi, Q. Wang, Q. Yin, Y. Wang, X. Hu, L. Wang, E. Liu, G. Xu, L. Lu, A. A. Soluyanov, H. Lei, Y. Shi, J. Luo, and Z.	Chen, Nat. Commun. \textbf{11}, 3985 (2020).
	
	\bibitem{Guguchia.NatComm}
	Z. Guguchia, J.~A.~T. Verezhak, D.~J. Gawryluk, S.~S. Tsirkin, J.-X. Yin, I. Belopolski, H. Zhou, G. Simutis, S.-S. Zhang, T.~A. Cochran, G. Chang, E. Pomjakushina, L. Keller, Z. Skrzeczkowska, Q. Wang, H.~C. Lei, R. Khasanov, A. Amato, S. Jia, T. Neupert, H. Luetkens, and M.~Z. Hasan, Nat. Commun. \textbf{11}, 559 (2020).
	
	
	\bibitem{Liu3}
	D. F. Liu, E. K. Liu, Q. N. Xu, J. L. Shen, Y. W. Li, D. Pei, A. J. Liang, P. Dudin, T. K. Kim, C. Cacho, Y. F. Xu, Y. Sun, L. X. Yang, Z. K. Liu, C. Felser, S. S. P. Parkin, Y. L. Chen, arXiv:2103.08113. 
	
	
	\bibitem{Shin}
	D.-H. Shin, J.-H. Jun, S.-E. Lee, M.-H. Jung, arXiv:2105.03892.
	
	
	\bibitem{Yanagi}
	Y. Yanagi, J. Ikeda, K. Fujiwara, K. Nomura, A. Tsukazaki, M.-T. Suzuki, Phys. Rev. B \textbf{103}, 205112 (2021).
	
	
	\bibitem{Savrasov}
	A. Rossi, V. Ivanov, S. Sreedhar, A. L. Gross, Z. Shen, E. Rotenberg, A. Bostwick, Ch. Jozwiak, V. Taufour, S. Y. Savrasov, I. M. Vishik,
	Phys. Rev. B \textbf{104}, 155115 (2021). 
	
	\bibitem{Koshino}
	M. Koshino and I. F. Hizbullah, Phys. Rev. B \textbf{93}, 045201 (2016).
	
	\bibitem{Armitage}
	N.P. Armitage, E. J. Mele, A. Vishwanath, Rev. Mod. Phys. \textbf{90}, 15001 (2018).
	
	\bibitem{Kassem}
	M. A. Kassem, PhD Dissertation (Kyoto Univ., 2016).
	
	\bibitem{Schnelle}
	W. Schnelle, A. Leithe-Jasper, H. Rosner, F. M. Schappacher, R. Poettgen, F. Pielnhofer, and R. Weihrich,
	Phys. Rev. B \textbf{88}, 144404 (2013).
	
	\bibitem{Volovik}
	G. E. Volovik,
	Phys. Usp. \textbf{61}, 89 (2018).
	
	\bibitem{Volovik1}
	G. E. Volovik, 
	Lect. Notes Phys. \textbf{718}, 31, 2007; arXiv:cond-mat 0601372.
	
	\bibitem{Volovik2}
	K. Zhang and G. E. Volovik, JETP Lett.\textbf{105},  519 (2017).
	
	
	\bibitem{Kane}
	C. L. Kane, P. A. Lee, N. Read,
	Phys. Rev. B\textbf{39}, 6880 (1989).
	
	\bibitem{IS} 
	V. Yu. Irkhin, Yu. N. Skryabin,	 JETP Lett. \textbf{106}, 167 (2017).
	
	\bibitem{FTT} V. Yu. Irkhin and M. I. Katsnelson,
	Sov. Phys. - Solid State \textbf{25}, 1947 (1983).
	
	\bibitem{JPCM}
	V. Yu. Irkhin and M.I. Katsnelson, J. Phys.: Cond. Mat. \textbf{2}, 7151 (1990).
	
	\bibitem{633a} P. W.  Anderson,
	Int. J. Mod. Phys. B\textbf{4}, 181 (1990).
	
	
	
	
	
	\bibitem{Senthil}
	T. Senthil,
	Phys. Rev. B \textbf{78}, 045109 (2008).
	
	\bibitem{Zarubin} 
	V. Yu. Irkhin, A. V. Zarubin, Phys. Rev. B \textbf{70}, 035116 (2004).
	
	
	\bibitem{Vojta} M. Vojta, Rep. Prog. Phys. \textbf{81}, 064501 (2018).
	
	
	\bibitem{Hasan} M. Z. Hasan, C. L. Kane, Rev. Mod. Phys. \textbf{82}, 3045 (2010).
	
	\bibitem{My}
	V. Yu. Irkhin, Yu. N. Skryabin, J. Exp. Theor. Phys. \textbf{133}, 116 (2021).
	
	\bibitem{Wan} 
	X. Wan, A. M. Turner, A. Vishwanath, and S. Y. Savrasov, Phys. Rev. B \textbf{83}, 205101 (2011). 
	
	\bibitem{Taguchi}
	Y. Taguchi, H. Oohara, N. Yoshizawa, Y. Nagaosa, and Y. Tokura, 
	Science \textbf{291}, 2573 (2001).
	
	\bibitem{Soh}
	J.-R. Soh, Ch. J. Yi, I. Zivkovic, N. Qureshi, A. Stunault, B. Ouladdiaf, J. A. Rodríguez-Velamazán, Y. Shi, A. T. Boothroyd, 	arXiv:2110.00475.
	
	\bibitem{Yang}
	T. Y. Yang, Q. Wan, Y. H. Wang, M. Song, J. Tang, Z. W. Wang, H. Z. Lv, N. C. Plumb, M. Radovic, G. W. Wang, G. Y. Wang, Z. Sun, R. Yu, M. Shi, Y. M. Xiong, N. Xu,
	arXiv:1906.07140 (2019).
	
	\bibitem{Alloul}
	I.F. Gilmutdinov, R. Schoenemann, D. Vignolles, C. Proust, I.R. Mukhamedshin, L. Balicas, H. Alloul, arXiv:2101.05252.
	
	
	\bibitem{Kono}
	T. Kono, M. Kakoki, T. Yoshikawa, X. Wang, K. Goto, T. Muro, R. Y. Umetsu, A. Kimura,
	Phys. Rev. Lett. \textbf{125}, 216403 (2020).
	
	
\end{thebibliography}
\end{document}